\documentstyle[12pt]{article}
\begin{document}
{\hbox to\hsize{November 1995 \hfill IASSNS-HEP 95/100}}\par
{\hbox to\hsize{hep-ph/9511446 \hfill BA-95-56}}\par
\begin{center}
{\LARGE \bf Large Neutrino Mixing Angles in  \\[0.1in]
Unified Theories}\footnote{Work supported in part by the
Department of Energy} \\[0.3in]
 {\bf K.S. Babu}\footnote{E-mail: babu@sns.ias.edu}\\[.05in]
{\it School of Natural Sciences\\
Institute for Advanced Study\\
Princeton, NJ 08540}\\[.15in]
and \\ [.1in]
{\bf S.M. Barr}\footnote{E-mail: smbarr@bartol.udel.edu} \\[.05in]
{\it Bartol Research Institute\\
University of Delaware\\
Newark, DE 19716}\\[.3in]
\end{center}

\begin{abstract}

Typically in unified theories the neutrino mixing angles,
like the Cabibbo-Kobayashi-Maskawa (CKM) angles of the quarks,
are related to the small mass ratios between fermions of
different generations and are therefore quite small.
A new approach for explaining the intergenerational
mass hierarchies is proposed here which, while giving small
CKM angles, naturally leads to neutrino angles of order unity.
Such large mixing angles may be required for a resolution of the
atmospheric neutrino anomaly and may also be relevant for the
solar neutrino puzzle.  The mechanism presented here provides
a framework in which novel approaches to the fermion mass
question can arise. In particular, within this framework
a variant of the texture idea allows highly predictive models to
be constructed, an illustrative example of which is given.
It is shown how the neutrino mixing angles may be
completely determined in such schemes.

\baselineskip = .30in

\end{abstract}
\newpage

\baselineskip = .30in

\section{Introduction}

There are hints of the possibility that certain neutrino mixing
angles may be large in contrast to the Cabibbo-Kobayashi-Maskawa
(CKM) mixing angles of the quarks. One such hint comes from atmospheric
neutrino data \cite{Kam}, an interpretation \cite{Gaisser}
of which suggests that the
mixing angle between $\nu_{\mu}$ and $\nu_{\tau}$ is of order
unity (sin$^22\theta_{\mu \tau} \simeq 0.5-1.0)$.
Another hint comes from the MSW explanation \cite{MSW} of the
solar neutrino problem \cite{Bahcall}, for which there
is a large angle solution (sin$^22\theta_{e\mu} \simeq 0.5-0.9)$
in addition to the small angle one for the mixing of the
$\nu_e$ \cite{Hata}.  Resolving the solar neutrino puzzle
via neutrino oscillation in vacuum also requires large
mixing:  sin$^22\theta_{e\mu} \simeq (0.70-1.0)$ \cite{Krastev}.

The trouble from a theoretical standpoint is that in the
context of unified theories, where there is the best hope
of understanding the neutrino mixing angles, these angles
are related to the hierarchy among the masses of the leptons,
just as the CKM angles are related to the hierarchy among the quark
masses, and hence are typically very small. In this paper
we propose a somewhat novel way to explain the intergenerational
mass hierarchies which has the interesting feature that it can
naturally give rise to large (order unity) neutrino mixings.
Moreover in the context of this mechanism it is possible to construct
highly predictive models of fermion masses
that give definite numbers for the neutrino mixing angles.
We give an illustrative example based on the
use of family symmetry and ``textures". However, as will be seen,
textures work in a somewhat different way in this framework
than in the familiar schemes.

\section{Large Neutrino Mixing Angles in Unified Theories}

In most schemes to explain the quark and lepton masses and mixings
the mass matrices have the typical form
\begin{equation}
U,D,L,N \sim \left( \begin{array}{ccc}
\epsilon_1 & \epsilon_1 & \epsilon_1 \\
\epsilon_1 & \epsilon_2 & \epsilon_2 \\
\epsilon_1 & \epsilon_2 & \epsilon_3
\end{array} \right)
\end{equation}

\noindent
where $\epsilon_1 \ll \epsilon_2 \ll \epsilon_3$.  Eq. (1) has to
be understood in the order of magnitude sense.  For the
general case, this clearly
gives a hierarchy among the masses $m_i/m_j \sim \epsilon_i/
\epsilon_j$ and angles $V_{ij} \sim \epsilon_i/\epsilon_j
\sim m_i/m_j$, for $i < j$. For the special case where
the (11) and (22) elements vanish \cite{Fritzsch},
as in many commonly considered ``textures", $V_{ij}
\sim \sqrt{m_i/m_j}$, which is still small.

In unified schemes, because of the relation between quarks and
leptons, the Dirac mass matrix of the neutrinos, $N$, is of the
same basic form as the mass matrices of the up-quarks, down-quarks,
and leptons, $U$, $D$, and $L$. The light neutrino mass matrix
is given by a ``see-saw" formula \cite{seesaw}
\begin{equation}
M_{\nu} = N^T M_R^{-1} N,
\end{equation}

\noindent
where $M_R$ is some superheavy matrix of masses for the right-handed
neutrinos. Unless $M_R$ itself exhibits some extreme ``hierarchical"
structure or has accidental cancellations \cite{Tanimoto},
it is clear from Eqs.
(1) and (2) that the neutrino mixing
angles are also typically given by
$V_{ij} \sim \epsilon_i/\epsilon_j \ll 1$, $i<j$.

It will now be shown that the hierarchy among generations can
arise as a consequence of mixing between the ordinary three
families and other fermions which have masses of order the
GUT scale.  We will call this ``exogenous mixing"
to distinguish it from mixing among the ordinary three families.
Such exogenous mixing can lead, as will be seen, to the structure
\begin{eqnarray}
U &=& H^T U_0 H, \nonumber \\
D &=& D_0 H, \nonumber \\
L &=& H^T L_0,\nonumber \\
N &=& N_0,
\end{eqnarray}
\noindent
where
\begin{equation}
H \simeq \left( \begin{array}{ccc}
\epsilon_1 & & \\
& \epsilon_2 & \\
& & \epsilon_3
\end{array} \right), \;\;\; \epsilon_1 \ll \epsilon_2 \ll
\epsilon_3,
\end{equation}

\noindent
and each of the matrices $U_0$, $D_0$, $L_0$, and $N_0$
has all its (nonvanishing) elements of the same order.
The hermitian matrix $H$ comes from the exogenous mixing
and gives rise to the intergenerational hierarchy of masses.
(The mass matrices in Eq. (3) are written so that
they are to be multiplied
on the right by the (left-handed) fermion fields and on the
left by the (left-handed) anti-fermion fields: $(f_L^c)_i M_{ij}
(f_L)_j$.)
This structure has two consequences. First, the mass
hierarchies go as
\begin{eqnarray}
m_d : m_s : m_b & \sim & m_e : m_{\mu} : m_{\tau}  \sim
\epsilon_1 : \epsilon_2 : \epsilon_3, \nonumber \\
m_u : m_c : m_t & \sim & \epsilon_1^2 : \epsilon_2^2 : \epsilon_3^2.
\end{eqnarray}

\noindent
This pattern receives some support from the data, as indeed the
hierarchy among the $u$, $c$, and $t$ masses is much greater
than that among the down quarks and leptons. For example,
$m_u/m_c \sim (m_d/m_s)^2$ and $m_c/m_t \sim (m_{\mu}/m_{\tau})^2$.

The second consequence of the structure in Eq. (3) is that
the mixing angles go as
\begin{eqnarray}
V_{ij}^{{\rm (quark)}} \equiv (V_{KM})_{ij} & \sim &
\epsilon_i/\epsilon_j,
\nonumber \\
V_{ij}^{({\rm lepton})} & \sim & 1.
\end{eqnarray}
The difference between the mixing of quarks and leptons is
clear from Eq. (3):  The hermitian matrix $H$ that contains
the hierarchy factors $\epsilon_{1,2,3}$ multiples the
left--handed quark fields yielding a hierarchy in the
quark mixing.  In the lepton sector, $H$ multiples
the right-handed charged lepton and therefore does not produce
a hierarchy in the left-handed lepton mixing.

\section{Exogenous Mixing}

To see how the form for $D$ in Eq. (3) arises, consider first the
case of one generation, $d$, which mixes ``exogenously" with
new fermions \cite{Bere}, $d'_L + \overline{d'_L}$, as follows:
\begin{equation}
{\cal L}_{{\rm d} \; {\rm mass}} = d^c_L D_0 d_L  +
\left( M \overline{d'_L} d'_L + m \overline{d'_L} d_L \right).
\end{equation}

\noindent
$M$ and $m$, being standard model singlets, are superlarge,
while $D_0$ which breaks $SU(2)_L$ is of weak scale.
The light and superheavy eigenstates are $d_L^{(l)}
= -\frac{m}{\sqrt{M^2 + m^2}} d'_L + \frac{M}{\sqrt{M^2 + m^2}}
d_L$ and $d_L^{(h)} = \frac{M}{\sqrt{M^2 + m^2}} d'_L +
\frac{m}{\sqrt{M^2 + m^2}} d_L$. Substituting, then,
$d_L = \frac{M}{\sqrt{M^2 + m^2}} d_L^{(l)} +
\frac{m}{\sqrt{M^2 + m^2}} d_L^{(h)}$ into the first term in Eq. (7)
gives
\begin{equation}
D = D_0 \left( \frac{M}{\sqrt{M^2 + m^2}} \right).
\end{equation}

\noindent The effect of exogenous mixing is to suppress the mass of
the light eigenstate, and for
$M \ll m$ it produces a hierarchy.

Now consider a simplified 3-generation case where each flavor
of down quark mixes as above: $d_i$ with $d'_i +
\overline{d'_i}$:
\begin{equation}
{\cal L}_{{\rm d} \; {\rm mass}} = \sum_{i,j = 1}^3 d^c_{L,i} D_{0,ij} d_{L,j}
+ \sum_{i=1}^3 \left( M_i \overline{d'_{L,i}} d'_{L,i} + m_i
\overline{d'_{L,i}} d_{L,i} \right).
\end{equation}

\noindent
As before, one can write for each generation
$d_{L,i} = \frac{M_i}{\sqrt{M_i^2 + m_i^2}} d_{L,i}^{(l)} +
\frac{m_i}{\sqrt{M_i^2 + m_i^2}} d_{L,i}^{(h)}$, and
so, defining $\epsilon_i \equiv \frac{M_i}{\sqrt{M_i^2 + m_i^2}}
= \left( 1 + \left( \frac{m_i}
{M_i} \right)^2 \right)^{-\frac{1}{2}}$, write

\begin{equation}
D= D_0 \left( \begin{array}{ccc}
\epsilon_1 & & \\ & \epsilon_2 & \\ & & \epsilon_3
\end{array} \right).
\end{equation}

\noindent
In the general case where the exogenous mixing is not
flavor-diagonal, but is described by arbitrary matrices
$M$ and $m$, one can easily show
\begin{equation}
D = D_0 H,
\end{equation}
where
\begin{equation}
H = \left( I + m M^{-1} M^{-1 \dag} m^{\dag} \right)^{-\frac{1}{2}}.
\end{equation}

Clearly, if there is exogenous mixing of fermions of type $f_L$ there will
appear a matrix $H$ on the right side of $F_0$, while exogenous mixing
of $f^c_L$ will produce such a matrix on the left side of $F_0$.
(Here and throughout $F_0$ stands for any of the matrices
$U_0$, $D_0$, $L_0$, or $N_0$.)
The pattern in Eq. (3) arises if there is mixing of all
the fermion types contained in a ${\bf 10}$ of SU(5), namely
$u^c$, $u$, $d$, and $l^+$. The matrices, $H$, appearing in all
these equations will be the same if this mixing is SU(5)-invariant.
For example, in an SO(10) model (one could, of course, do the
same kind of thing in an SU(5) model) there could be in addition
to the three ordinary families, ${\bf 16}_i$, three adjoints
of fermions, ${\bf 45}_I$ (actually one only requires two)
with mass terms
\begin{equation}
{\cal L}_{{\rm mass}} = \sum_{i,j} {\bf 16}_i F_{0,ij} {\bf 16}_j
+ \left( \sum_{i,I} e_i^I {\bf 16}_i {\bf 45}_I \langle
\overline{{\bf 16}}_H \rangle + \sum_{I,J} M_{IJ} {\bf 45}_I
{\bf 45}_J \right).
\end{equation}

\noindent
The matrix $F_0$ is considered to come from some unspecified
Yukawa terms whose structure is complicated enough to
incorporate SO(10) and SU(5) breaking so as to give different
submatrices $U_0$, $D_0$, $L_0$, and $N_0$. Each of these
matrices is assumed to have no significant hierarchy among
its eigenvalues, the intergenerational hierarchy coming
entirely from the matrix $H$, which from Eq. (12) is given by
the expression
\begin{equation}
H = \left( I + v_R^2 e M^{-1} M^{-1 \dag} e^{\dag} \right)^{-\frac{1}
{2}}.
\end{equation}

\noindent
Here $e$ is a matrix with elements given by $e_i^I$ and
$v_R$ is the unification-scale vacuum expectation value
of the SU(5)-singlet component of the $\overline{{\bf 16}}_H$.
This singlet vacuum expectation value couples the ${\bf 10(16_i)}$
to the ${\bf \overline{10}(45_I)}$. It also couples the ${\bf 1(16_i)}$
to the ${\bf 1(45_I)}$ so that a mixing matrix analogous to $H$
will appear for the left-handed antineutrinos on the left side
of $N_0$. More precisely, it is easily shown that the mass
matrix of the light left-handed neutrinos after integrating
out the superheavy neutrinos is
\begin{equation}
M_{\nu} = \frac{4}{5} v_R^{-2} \left( N_0^T e^{-1 T} M e^{-1} N_0
\right).
\end{equation}

\noindent
The factor of $\frac{4}{5}$ is an SO(10) Clebsch-Gordon coefficient.
For $H$ to have the hierarchical structure shown in Eq. (4) there
needs to be a hierarchy either (case A) among the eigenvalues of the
Yukawa coupling matrix, $e_i^I$, or (case B) among the eigenvalues
of the mass matrix, $M_{IJ}$. (Or in both matrices; but we shall
primarily
focus on the two cases mentioned as being conceptually simpler and
more plausible.) If the hierarchy is in the eigenvalues of $e_i^I$
rather than $M_{IJ}$ then one expects from Eq. (14) that the
eigenvalues
of $H$, which were denoted above $\epsilon_1$, $\epsilon_2$,
and $\epsilon_3$, that is the hierarchy factors, are given by
\begin{equation}
\epsilon_i \sim \left( 1 + \left( \frac{v_R}{M} \right)^2 (e_i)^2
\right)^{-\frac{1}{2}},
\end{equation}

\noindent
where $e_i$ is the ${\rm i}^{{\rm th}}$ eigenvalue of $e_i^I$, and $M$
is a typical eigenvalue of $M_{IJ}$. These $\epsilon_i$ are all less than
or equal to one.  Only the ratios $\epsilon_1/\epsilon_3$ and
$\epsilon_2/\epsilon_3$ are relevant for fermion masses, so one can set
$\epsilon_3 = 1$ if desired.  We shall however
keep the $\epsilon_3$ dependence explicit.
In simple SO(10)
schemes with a single ${\bf 10}$-plet of Higgs that generates the top
and the bottom masses $\epsilon_3$
cannot be much less than unity
because from Eq. (3) one has $\tan \beta \sim
\epsilon_3^{-1} m_t/m_b$.  That is, $(v_R/M) e_3
\stackrel{_<}{_\sim} 1$.  (In the special case where there are only
two ${\bf 45}_I$ of fermions, $\epsilon_3=1$.)
The hierarchy among the $\epsilon_i$ is then given approximately
by
\begin{equation}
\epsilon_1 : \epsilon_2 : \epsilon_3 \sim \left( \frac{v_R}{M} e_1
\right)^{-1} : \left( \frac{v_R}{M} e_2 \right)^{-1} : 1.
\end{equation}

\noindent
In contrast, the hierarchy among the neutrino masses, by Eq. (15),
goes as $\left( \frac{v_R}{M} e_1 \right)^{-2} : \left( \frac{v_R}{M}
e_2 \right)^{-2} : \left( \frac{v_R}{M} e_3 \right)^{-2}$.
That is, for ``case A"
\begin{eqnarray}
m_{\nu_e}/m_{\nu_{\mu}} & \sim & (m_e/m_{\mu})^2 \sim (m_d/m_s)^2
\sim m_u/m_c, \nonumber \\
m_{\nu_{\mu}}/m_{\nu_{\tau}} & \stackrel{_<}{_\sim} &
(m_{\mu}/m_{\tau})^2 \sim (m_s/m_b)^2 \sim m_c/m_t.
\end{eqnarray}
\noindent
In case B, where the hierarchy is in the eigenvalues of $M_{IJ}$,
one finds from Eq. (15)
\begin{eqnarray}
m_{\nu_e}/m_{\nu_{\mu}} & \sim & m_e/m_{\mu} \sim m_d/m_s
\sim \sqrt{m_u/m_c}, \nonumber \\
m_{\nu_{\mu}}/m_{\nu_{\tau}} & \stackrel{_<}{_\sim} &
m_{\mu}/m_{\tau} \sim m_s/m_b \sim \sqrt{m_c/m_t}.
\end{eqnarray}

\noindent
Notice that in both cases the neutrino mass hierarchy is weaker than
the often assumed see-saw result $m_{\nu_e}: m_{\nu_{\mu}} :
m_{\nu_{\tau}} \sim m_u^2 : m_c^2 : m_t^2$.
Also, in both cases
the neutrino mixing angles are {\it of order unity}.

Two remarks about the mass hierarchy shown in Eqs. (18)-(19) are in
order.  If the solar neutrino puzzle is solved via
$(\nu_e-\nu_\mu)$ MSW
oscillations, the
mass of $\nu_\tau$ will be typically
in the range $m_{\nu_\tau} \sim (0.1-1)~eV$
(using $m_{\nu_\mu} \sim 3 \times 10^{-3}~eV$).
This is in the right range for solving the atmospheric neutrino
anomaly via $(\nu_\mu-\nu_\tau)$ oscillations.  It is also worth
noting that the usual two-flavor analysis of accelerator and
reactor neutrino data will not be sufficient here since
all three mixing
angles are large, a three flavor analysis as was done
in Ref. \cite{Lisi}
with one dominant mass \cite{BPW}  will be required.

\section{Predictive Schemes}

Beyond the {\it qualitative} results on the hierarchies and mixing angles
it is possible to construct predictive schemes within this framework.
Suppose that there is some family symmetry (which could be discrete)
which acts on the index $i$ of the three ordinary families, ${\bf 16}_i$,
but not on the index $I$ of the extra fields, ${\bf 45}_I$, with which they
mix.
Such a symmetry will generally produce some definite ``textures"
for $U_0$, $D_0$, $L_0$, and $N_0$. It also distinguishes the
quantities $\left| \vec{e}_i \right| \equiv \sqrt{\sum_I (e_i^I)^2}$
and allows in a natural
way a hierarchy to exist among them: $\left| \vec{e}_1 \right| \gg
\left| \vec{e}_2 \right| \gg \left| \vec{e}_3 \right|$. We assume such
a hierarchy. On the other
hand the matrix $M_{IJ}$ is unconstrained by such a family symmetry
and will be taken to be an arbitrary matrix exhibiting no particular
hierarchy. Thus this scenario falls under case A.

Under these assumptions, the matrix $H$ given by Eq. (14) is not
diagonal.
However, it can be written $H = U^{\dag} \overline{H} U$, where
$\overline{H} = {\rm diag}( \epsilon_1, \epsilon_2, \epsilon_3)$,
and $U \simeq I + i \theta$ is a unitary matrix with $\theta_{ij}
\sim \epsilon_i/\epsilon_j \ll 1$, ($i < j$). It can then easily be
shown
from Eq. (3) that the expressions for the quark and charged-lepton
mass ratios and for the Cabibbo-Kobayashi-Maskawa angles do not
depend on the
angles $\theta_{ij}$ to leading order in the small quantities $\epsilon_i/
\epsilon_j$, $(i < j)$. Thus, the ``exogenous mixing" effectively
introduces only
{\it two} parameters for describing the charged-fermion sector to this
order, namely $|\epsilon_1/\epsilon_3 |$ and
$| \epsilon_2/\epsilon_3 |$.

If one defines the vectors $\vec{D}_i$, $\vec{L}_i$, and $\vec{N}_i$
by $(\vec{D}_i)_j \equiv D_{0,ji}$, $(\vec{L}_i)_j \equiv L_{0,ij}$,
and $(\vec{N}_i)_j \equiv N_{0,ij}$, then to leading order in
the hierarchy factors $| \epsilon_i/\epsilon_j |$ one can write
\begin{equation}
\begin{array}{ccl}
m_b & = & \left| \epsilon_3 \right| \left| \vec{D}_3 \right|, \\
m_s & = & \left| \epsilon_2 \right| \left| \vec{D}_2 \times \vec{D}_3 \right|/
\left| \vec{D}_3 \right|, \\
m_d & = & \left| \epsilon_1 \right| \left| \vec{D}_1 \cdot \vec{D}_2
\times \vec{D}_3 \right| / \left| \vec{D}_2 \times \vec{D}_3 \right|,
\end{array}
\end{equation}

\begin{equation}
\begin{array}{ccl}
m_{\tau} & = & \left| \epsilon_3 \right|  \left| \vec{L}_3 \right|, \\
m_{\mu} & = & \left| \epsilon_2 \right| \left| \vec{L}_2 \times \vec{L}_3
\right|/\left| \vec{L}_3 \right|, \\
m_e & = & \left| \epsilon_1 \right| \left| \vec{L}_1 \cdot \vec{L}_2
\times \vec{L}_3 \right|/ \left| \vec{L}_2 \times \vec{L}_3 \right|,
\end{array}
\end{equation}

\noindent
while on the other hand
\begin{equation}
\begin{array}{ccl}
m_t & = & \left| \epsilon_3 \right|^2 U_{0,33} \\
m_c & = & \left| \epsilon_2 \right|^2 \det_{23} U_0 / U_{0,33} \\
m_u & = & \left| \epsilon_1 \right|^2 \det U_0 / \det_{23} U_0.
\end{array}
\end{equation}

\noindent
To leading order the Kobayashi-Maskawa elements are given by
\begin{equation}
\begin{array}{ccl}
V_{cb} & = & \left| \frac{\epsilon_2}{\epsilon_3} \right|
\left[ \frac{\vec{D}_2 \cdot \vec{D}_3}{(\vec{D}_3)^2} -
\frac{U_{0,32}}{U_{0,33}} \right], \\
V_{us} & = & \left| \frac{\epsilon_1}{\epsilon_2} \right|
\left[ \frac{\vec{D}_1 \cdot \vec{D}_2 - \vec{D}_1 \cdot
\hat{D}_3 \vec{D}_2 \cdot \hat{D}_3}{\mid \vec{D}_2 \times
\hat{D}_3 \mid^2} - \frac{U_{0,33} U_{0,21} - U_{0,31} U_{0,23}}
{\det_{23} U_0} \right], \\
V_{ub} & = & \left| \frac{\epsilon_1}{\epsilon_3} \right|
\left[ \frac{\vec{D}_1 \cdot \vec{D}_3}{(\vec{D}_3)^2}
- \frac{U_{0,31}}{U_{0,33}} +
\left( \frac{U_{0,33} U_{0,21} - U_{0,31} U_{0,23}}
{\det_{23} U_0} \right) \left( \frac{\vec{D}_2
\cdot \vec{D}_3}{(\vec{D}_3)^2} -
\frac{U_{0,32}}{U_{0,33}} \right) \right].
\end{array}
\end{equation}

{}From the fact that the neutrino mixing angles are of order unity
it is obvious that they cannot depend (in leading order) on
the small parameters $\epsilon_i$ and $\theta_{ij}$ that
characterize the matrix $H$, and thus the neutrino angles
must be computable in terms of the elements of $L_0$ and $N_0$.

First consider $L$. By Eqs. (3), (13), and (14) this is given by
$L = H L_0 \cong \left( \begin{array}{ccc} \epsilon_1 & & \\
& \epsilon_2 & \\ & & \epsilon_3 \end{array} \right) L_0
= \left( \begin{array}{c} \epsilon_1 \vec{L}_1^T \\
\epsilon_2 \vec{L}_2^T \\ \epsilon_3 \vec{L}_3^T \end{array}
\right)$. Here the angles $\theta_{ij}$, which only come in at
higher order, have been neglected; so that $H$ is effectively
diagonal. It is clear that if $L$ is diagonalized by
$\overline{L} = V_L^{\prime \dag} L V_L$, then
\begin{equation}
V_L \cong \left( \frac{\vec{L}_2 \times \hat{L}_3}{\mid \vec{L}_2
\times \hat{L}_3 \mid}, \frac{\vec{L}_2 - \vec{L}_2 \cdot \hat{L}_3
\hat{L}_3}{\mid \vec{L}_2 \times \hat{L}_3 \mid}, \hat{L}_3 \right).
\end{equation}

\noindent ($\hat{L}_3$ has unit norm.)
Moreover, one can write $M_{\nu} = N_0^T H^{\prime T} H' N_0$
(see Eq. (15)), where $H'$ has a form similar to that of $H$.
Thus, one can show that to leading order the matrix $V_N$
that diagonalizes $M_{\nu}$ is given by
\begin{equation}
V_N \cong \left( \frac{\vec{N}_2 \times \hat{N}_3}{\mid \vec{N}_2
\times \hat{N}_3 \mid}, \frac{\vec{N}_2 - \vec{N}_2 \cdot \hat{N}_3
\hat{N}_3}{\mid \vec{N}_2 \times \hat{N}_3 \mid}, \hat{N}_3 \right).
\end{equation}

\noindent
And the neutrino mixing matrix is given by
\begin{equation}
V_{{\rm lepton}} = V_N^{\dag} V_L.
\end{equation}

\section{An Example}

To make these ideas concrete, we present an example of a texture
that well reproduces the pattern of quark and lepton masses and gives
definite predictions for the neutrino mixing angles, which are large.
Let the matrices $U_0$, $D_0$, $L_0$, and $N_0$ have the form

\begin{equation}
F_0 = \left( \begin{array}{ccc}
0 & C X_f & 0 \\
C X_{f^c} & D (B-L)_f & B (I_{3R})_f \\
0 & B (I_{3R})_{f^c} & A
\end{array} \right) v^{(\prime)}.
\end{equation}

\noindent
Here $X$, $B-L$, and $I_{3R}$ are SO(10) generators. The subscript
($f$ or $f^c$) indicates whether it is the charge of the fermion
or antifermion. Such a form can arise from an SO(10) model.
($A$ arises from a Higgs {\bf 10}-plet, $B$ and $C$ from
effective ${\bf
10} \times {\bf 45}$ operators with $\left
\langle {\bf 45} \right \rangle
\propto I_{3R}$ and $X$ respectively \cite{Barr}, and $D$
from an effective $\overline{\bf 126}$ \cite{BM}.)
Specifically the mass matrices are given by
\begin{eqnarray}
U_0 &=& \left( \matrix{
0 & C & 0 \cr C & D/3 & 0 \cr 0 & -B/2 & A}
\right) v,  ~~D_0 = \left( \matrix{
0 & C & 0 \cr -3C & D/3 & 0 \cr 0 & B/2 & A}
\right) v',  \nonumber \\
L_0 &=& \left( \matrix{
0 & -3C & 0 \cr C & -D & 0 \cr 0 & B/2 & A}
\right) v',  ~~N_0 = \left(\matrix{
0 & -3C & 0 \cr 5C & -D & 0 \cr 0 & -B/2 & A}
\right) v.
\end{eqnarray}

If $\left| \epsilon_2/\epsilon_3 \right| = 0.08$,
$\left| \epsilon_1/\epsilon_3 \right| = 0.02$,
$B/A = 0.4$, $C/A = 0.06$, and $D/A = 0.75$, one gets an
excellent fit: $m_{\tau}/m_b = 1.02$,
$m_{\mu}/m_s \cong 3.0$, $m_e/m_d \cong 0.33$,
$m_{\mu}/m_{\tau} \cong 0.06$, $m_e/m_{\mu} \cong
5 \times 10^{-3}$, $m_c/m_t \cong 1.6 \times 10^{-3}$,
$m_u/m_c \cong 3.5 \times 10^{-3}$, $V_{us} \cong 0.22$,
$V_{ub} \cong 0.002$, and $V_{cb} \cong 0.03$. (These quantities
are all defined at the unification scale.)

Note that this texture model has six parameters ($v/v'$,
$B/A$, $C/A$, $D/A$, $\left| \epsilon_2/\epsilon_3 \right|$,
and $\left| \epsilon_1/\epsilon_3 \right|$) to fit
eleven quantities (eight mass ratios and three CKM angles,
ignoring CP violation.  CP violation can arise in the
model without increasing parameters if $B$ in Eq. (27) is
proportional to $I_{3R}$ approximately rather than exactly.  It
can also arise through the new phases of the
supersymmetry breaking sector as discussed
in Ref. \cite{Dann}).   Compared to usual texture models \cite{Raby}
this kind of model requires two extra parameters to give
the hierarchies, namely the $\left| \epsilon_i/\epsilon_j \right|$,
but economizes on parameters because the same form of texture
can be used for $U_0$ and $D_0$. (In conventional texture
models this would cause $V_{cb}$ to be too big.) Moreover,
one has as well definite predictions for the neutrino angles:
\begin{equation}
V_{{\rm lepton}} \cong \left( \begin{array}{ccc}
0.95 & 0.3 & -0.088 \\
-0.3 & 0.87 & -0.39 \\
0.032 & 0.4 & 0.92 \end{array} \right).
\end{equation}

One sees that in this example the neutrino angles are
significantly bigger than their Cabibbo-Kobayashi-Maskawa
counterparts; for example the (13) component is about
40 times larger and the (23) component is about 10 times
or more larger. Still, the neutrino angles are somewhat
smaller than unity. This can be traced to the smallness
of the parameter $C/A$. (Even normalizing the generator
$X$ so that ${\rm tr}_{16} \tilde{X}^2 = 2$ one still has the
ratio $\tilde{C}/A = 0.14$.) In general, in models of this
type where the ratios of parameters in the mass matrices
$F_0$ are close to unity, so also will be the neutrino angles.

\section{Conclusion}

We have shown that ``exogenous mixing"
between the ordinary families and new fermions
can explain the hierarchy among the generations in a
way that naturally gives neutrino mixing angles of order
unity while at the same time giving small Cabibbo-Kobayashi-Maskawa
angles. We have explored in this letter one approach
using this idea. There are other possibilities. For example,
in the ``short version" of the SO(10) model presented
in Ref. ({\cite{BB})
there is exogenous mixing of the fermions in the $\overline
{{\bf 5}}({\bf 16}_i)$ with those in SO(10) ${\bf 10}$'s
which tends (in that model) to make $V_{\nu_\mu \tau}$
and $V_{\nu_\tau \mu}$ large.
Indeed it was consideration of that example which inspired
the present work.

The approach described in this paper, summarized in Eqs. (3)-(4),
leads to a larger hierarchy among the charge-$\frac{2}{3}$
quarks than among the charge-$\frac{1}{3}$ quarks or charged leptons.
This is what is actually seen in the data.
Within this framework, as
shown in the explicit example presented, it is possible to construct
highly predictive models that give definite and large values
of the neutrino mixing angles.

\end{document}